\documentclass[pra,aps,superscriptaddress,showpacs,tightenlines,amsmath]{revtex4}
\usepackage{epsfig}
\usepackage{amsmath,amssymb}

\begin{document}

\title{Teleportation algorithm using two species of entangled pairs}

\author{Francisco Delgado}
\email{fdelgado@itesm.mx}
\affiliation{Mathematics and Physics Department, Quantum Information Processing Group, Tecnologico de Monterrey, Campus Estado de Mexico, Atizapan, Estado de Mexico, CP. 52926, Mexico}

\date{\today}
\begin{abstract}
Teleportation algorithm assumes specific Bell states as input, but actual sources typically generates more than one. This work presents a teleportation algorithm for a two Bell states mixture, including remaining distortion from previous control process. 

\pacs{03.67.Ac; 03.67.Hk; 03.67.-a}

\end{abstract}

\maketitle

\section{Introduction}
Teleportation algorithm is a central result not only in Quantum Engineering for information transfer, but in whole Quantum Mechanics. Nevertheless, this algorithm is constructed assuming a specific type of Bell state as input, when in practice, the sources of entangled pairs tipically generates more than one kind of them (specially spin entangled ones based on electronic gases and solid state spin ensembles as is reported in \cite{saraga1, saraga2, morton1, simmons1}).  

Some attempts to teleport specific states are been reported \cite{kim1, chen1}. This work presents an alternative general teleportation algorithm for one qubit which works with two standard Bell states as statistical mixture. The process works for light and spin technology, but for spin resources, it includes an additional gain for some aspects related with the possible remaining distortion obtained from previous control process on the pairs avoiding self-distortion generated by because of parasite magnetic fields \cite{delgado1}.

\section{Description of input states}
The problem stated here is related with an emission source of entangled pairs (typically $\left| \beta_{01} \right>$ and $\left| \beta_{10} \right>$; other types can be first transformed in these by using adequate initial transformations as $X$ and $Z$), which generates a statistical mixture of Bell states. This source is intendeed to use as teleportation resource for the traditional algorithm in Figure 1 \cite{nielsen1}, which can generate this effect departing from $\left| \beta_{10} \right>$ Bell state, in order to teleport the $\left| \psi' \right>=\alpha \left| 0 \right> + \beta \left| 1 \right>$ state.

Nevertheless, the presence of two Bell species as resources affects the process, inclusively without consider noise or distortion phenomenon caused by Heisenberg interaction for spin entangled pairs. In the discussion we will use the generic state:

\begin{equation}
\left| \beta \right>)=\sin \theta \left| \beta_{01} \right> - \cos \theta \left| \beta_{10} \right>
\end{equation}

\noindent which includes as particular cases $\left| \beta_{01} \right>$ and $\left| \beta_{10} \right>$ states for $\theta=0, \pi/2$ respectively. In addition, for spin systems we can include the remaining distortion of control process of Heisenberg interaction: $H= -J \vec{\mathbf{\sigma}}_1 \cdot \vec{\mathbf{\sigma}}_2+B_1 {\sigma_1}_z +B_2 {\sigma_2}_z$ by some time $t$ \cite{delgado1}. With this, the last state becomes:

\begin{equation}
\left| \beta \right>)=\sin \theta \left| \beta_{01} \right> - e^{2 i \pi n \delta} \cos{2 \pi n \delta} \cos \theta \left| \beta_{10} \right> + i e^{2 i \pi n \delta} \sin{2 \pi n \delta} \cos \theta \left|\beta_{00} \right>
\end{equation}

\noindent where $\delta=j-Q(j)$, being $Q(j)$ a rational approximation of $j=J/(B_{-}^2+4J^2)^{1/2},B_{-}=B_1-B_2$ in agreement with \cite{delgado1}. Note that this possible remaining distortion can be generated by limited knowledge of $j$ or $B_{-}$ (in this case $\delta \approx -B_{-}^2/4J^2$), but otherwise it can be deliberately generated to have control parameters in further processes which will use the state.

\section{Modified teleportation algorithm: outcomes and fidelity} 
Because both, distortion and input of two possible species destroy the outcome of the process, one can suppose as first instance, that eliminating the final correction of traditional algorithm, $X^{M_2}Z^{M_1+1}$, we could improve the outcome. Direct calculations show that in this case the fidelity is 0.5 for any value of $\theta$ parameter and gets even worse when distortion is present, $n \delta \ne 0$ (Figure 2a). 

If we leave the algorithm without change, we obtain the teleported states $\left|\psi'' \right>$ showed in Table 1 for each measurement outcome. It gives a fidelity equal to zero for $\theta=\pi/2$ (it means, for the $\left| \beta_{10} \right>$ specie), independently of distortion (Figure 2b). 

\begin{table}[htb]
 \centering \caption{Application of traditional and modified teleportation algorithm for $\left|\psi' \right> \otimes \left|\beta \right>$.}
\begin{tabular}{cl}
    \hline
    Measurement & Teleportated state  \\
    \hline
    $\left|00 \right>$ & $\left|\psi'' \right>=-\frac{1}{2}(\alpha \cos \theta - \beta \sin \theta )\left| 0 \right>  -\frac{1}{2}(\beta e^{4 i \pi n \delta} \cos \theta + \alpha \sin \theta )\left| 1 \right>$ \\
    & $\left|\psi''' \right>=-\frac{1}{2}(\alpha + i \beta e^{2 \pi n \delta} \sin 2 \pi n \delta \sin 2\theta )\left| 0 \right>  -\frac{1}{2} \beta (e^{4 i \pi n \delta} \cos^2 \theta + \sin^2 \theta )\left| 1 \right>$ \\
      \hline
    $\left|01 \right>$ & $\left|\psi'' \right>=-\frac{1}{2}(\alpha e^{4 i \pi n \delta} \cos \theta + \beta \sin \theta )\left| 0 \right>  -\frac{1}{2}(\beta \cos \theta - \alpha \sin \theta )\left| 1 \right>$ \\
    & $\left|\psi''' \right>=-\frac{1}{2}(\alpha (e^{4 i \pi n \delta} \cos^2 \theta-\sin^2 \theta) + \beta \sin 2\theta )\left| 0 \right>  -\frac{1}{2}(\beta \cos 2\theta - \alpha e^{2 \pi n \delta} \cos 2 \pi n \delta \sin 2\theta )\left| 1 \right>$ \\
    \hline
    $\left|10 \right>$ & $\left|\psi'' \right>=-\frac{1}{2}(\alpha \cos \theta + \beta \sin \theta )\left| 0 \right>  -\frac{1}{2}(\beta e^{4 i \pi n \delta} \cos \theta - \alpha \sin \theta )\left| 1 \right>$ \\
    & $\left|\psi''' \right>=-\frac{1}{2}(\alpha \cos 2\theta + \beta e^{2 \pi n \delta} \cos 2 \pi n \delta \sin 2\theta ) 
    \left| 0 \right>  -\frac{1}{2} (\beta (e^{4 i \pi n \delta} \cos^2 \theta-\sin^2 \theta) - \alpha \sin 2\theta ) \left| 1 \right>$ \\
    \hline
    $\left|11 \right>$ & $\left|\psi'' \right>=\frac{1}{2}(\alpha e^{4 i \pi n \delta} \cos \theta - \beta \sin \theta )\left| 0 \right>  +\frac{1}{2}(\beta \cos \theta + \alpha \sin \theta )\left| 1 \right>$ \\
    & $\left|\psi''' \right>=\frac{1}{2}\alpha (e^{4 i \pi n \delta} \cos^2 \theta + \sin^2 \theta )\left| 0 \right>  + \frac{1}{2} (\beta - i \alpha e^{2 \pi n \delta} \sin 2 \pi n \delta \sin 2\theta ) \left| 1 \right>$ \\
    \hline
   \end{tabular}
    \end{table}

But an adequate analysis of post-measurement states $\left|\psi'' \right>$ in Table 1 suggest some convenient correction: to include a $Y$-rotation for an angle equals to $\theta$. By introducing the following gates at the end of the traditional process on $\left|\psi'' \right>$:

\begin{equation}
H U_Y(\theta) H \equiv H (\cos \theta - i \sin \theta Y) H
\end{equation}

\noindent we obtain the $\left|\psi''' \right>$ states included in Table 1 too. The $U_Y(\theta)$ gate is easily obtained with polarimeters for light and with transverse magnetic fields for spins. Three main results are observed: for $n \delta=0$, all teleported states are exactly $\left|\psi' \right>$ for the cases $\theta=0, \pi/2$; for these values of $\theta$, the probability for each mesurement is in all cases $1/4$ as in the traditional process; and, still for  $n \delta \ne 0$, the state with $\theta=\pi/2$ is exactly $\left|\psi' \right>$, but for $\theta=0$ the teleported states are slightly distorted by a phase factor in one of both terms. Figure 2c shows some details in terms of fidelity. In the whole graphics of Figure 2, fidelity is presented as function of $\theta$, the $\left|\psi' \right>$ state by means of $\alpha_1 \in \mathbb{R}$ and the strenght of distortion, $n \delta$.

Resuming the findings, note in Figure 2a how the fidelity is not better than 0.5 (it is equal to 0.5 when $n \delta=0$). In Figure 2b, note how fidelity is zero for $\theta=\pi/2$ and how for $theta=0$ it drops to zero for greater distortion. Finally, note in Figure 2c how fidelity it is one for $\theta=0, \pi/2$ when $n \delta=0$ and reduced only for $\theta=0$ when distortion is present.

\section{Conclusions} 
The quantum algorithm presented works efficiently for the two Bell species considered, being limited only for the possibility to stablish adequate previous control of the original resource states produced by noise or distortion (in both cases, light polarization or spin). In the case of spin entangled particles, the presence of distortion can be eliminated only by knowing the exact strength of it, but limitated distortion still gives sufficient fidelity to obtain a faithful teleportated state (aproximately linear for the first 5\% of $n \delta$, giving in such limit case until 95\% of fidelity).

Future developments in this direction should to consider more complex teleportation process (by example, multipartite teleportation) and to introduce additional control parameters to include more Bell states in the statistical source or selective fidelity control for some specific species.

\section*{FIGURE CAPTIONS}
\subsection*{Figure 1}
Quantum teleportation algorithm for one qubit and modification to adapt it for two species of entangled states as resource.

\subsection*{Figure 2}
Fidelity, $F$, for teleportation algorithm using the $\left|\psi' \right> \otimes \left|\beta \right>$ state depicted in the text: a) without $X^{M_2}Z^{M_1+1}$-correction and $H U_Y(\theta) H$-rotation, b) with correction and without rotation, and c) with correction and rotation. Dark surface shows the $n \delta=0$ case without distortion and the light surfaces are the fidelities for different cases of $n \delta \ne 0$, ranging from 0.05 to 0.25 (for the worst distortion and fidelity).

\end{document}